\documentclass[preprint,tightenlines,showpacs,amsmath,amssymb]{revtex4}

\usepackage{epsfig,epsf,graphics,psfrag}
\usepackage{times}
\usepackage{float}
\usepackage{color}
\usepackage{amsfonts,amssymb,stmaryrd,latexsym,amsmath}
\usepackage{multirow}
\usepackage{array} 
\usepackage{hhline}
\usepackage{booktabs}
\usepackage{enumerate}
\usepackage{slashed}

\begin{document}

\bibliographystyle{unsrt}

\title{How to understand the underlying structures of $X(4140)$, $X(4274)$, $X(4500)$ and $X(4700)$}

\author{Xiao-Hai Liu$^1$\footnote{liuxh@th.phys.titech.ac.jp} }

\affiliation{ $^1$Department of Physics, H-27, Tokyo Institute of Technology, Meguro, Tokyo 152-8551, Japan}

\date{\today}

\begin{abstract}

We investigate the possible rescattering effects which may contribute to the process $B^+\to J/\psi\phi K^+$. It is shown that the $D_{s}^{*+}D_{s}^{-}$ rescattering via the open-charmed meson loops, and $\psi^\prime \phi$ rescattering via the $\psi^\prime K_1$ loops may simulate the structures of $X(4140)$ and $X(4700)$, respectively. However, if the quantum numbers of $X(4274)$ ($X(4500)$) are $1^{++}$ ($0^{++}$), it is hard to  to ascribe the observation of $X(4274)$ and $X(4500)$ to the $P$-wave threshold rescattering effects, which implies that $X(4274)$ and $X(4500)$ could be genuine resonances. We also suggest that $X(4274)$ may be the conventional orbitally excited state $\chi_{c1}(3P)$.




\end{abstract}

\maketitle

\section{Introduction}

Very recently, the LHCb collaboration reported the observation of several resonance-like structures in $J/\psi \phi$ invariant mass distributions in $B^+\to J/\psi\phi K^+$ decays \cite{Aaij:2016iza,Aaij:2016nsc}. Their masses, widths and favourable quantum numbers are
\begin{eqnarray}\label{Xstates}
&&	M_{X(4140)}=4146.5\pm 4.5^{+4.6}_{-2.8}\ \mbox{MeV},\ 	\Gamma_{X(4140)}=83\pm 21^{+21}_{-14}\ \mbox{MeV},\ 
J^{PC}=1^{++},  \nonumber \\
&&	M_{X(4274)}=4273.3\pm 8.3^{+17.2}_{-3.6}\ \mbox{MeV},\ 	\Gamma_{X(4274)}=56\pm 11^{+8}_{-11}\ \mbox{MeV},\ 
J^{PC}=1^{++},  \nonumber \\
&&	M_{X(4500)}=4506\pm 11^{+12}_{-15}\ \mbox{MeV},\ 	\Gamma_{X(4500)}=92\pm 21^{+21}_{-20}\ \mbox{MeV},\ 
J^{PC}=0^{++},  \nonumber \\
&&	M_{X(4700)}=4704\pm 10^{+14}_{-24}\ \mbox{MeV},\ 	\Gamma_{X(4700)}=120\pm 31^{+42}_{-33}\ \mbox{MeV},\ 
J^{PC}=0^{++},
\end{eqnarray}
among which the higher states $X(4500)$ and $X(4700)$ are firstly reported by the LHCb collaboration. 
$Y(4140)$ and $Y(4274)$ were firstly observed by the CDF collaboration in the $J/\psi\phi$ invariant mass distribution from $B\to K J/\psi\phi$ decays \cite{Aaltonen:2009tz,Aaltonen:2011at}. The presence of $Y(4140)$ in $B$ decays was later confirmed by the CMS and D0 collaborations~\cite{Chatrchyan:2013dma,Abazov:2013xda,Abazov:2015sxa}. Another state $X(4350)$ was reported by the Belle collaboration from the two photon process $\gamma\gamma$$\to$$J/\psi\phi$ \cite{Shen:2009vs}. $Y(4140)$ and $Y(4274)$ were also expected to be produced in the two photon fusion reaction, but neither of them was observed \cite{Shen:2009vs}.

These resonance-like peaks in the $J/\psi\phi$ invariant mass spectrum are very intriguing, because they may contain both a $c\bar{c}$ pair and and an $s\bar{s}$ pair, which implies that these states may be exotic. Taking into account their masses and decay modes, there are some suggestions that $Y(4140)$ and $Y(4274)$ are probably the hadronic bound states of $D_s^{*+}D_s^{*-}$ and $D_{s0}^{+}D_s^{-}$, respectively \cite{Liu:2009ei,Shen:2010ky,Liu:2010hf,He:2011ed,Finazzo:2011he,Wang:2011uk,Wang:2009wk,Wang:2014gwa,Albuquerque:2009ak,Ma:2014ofa,Ma:2014zva}. The tetraquark state $c\bar{c}s\bar{s}$ is also one popular explanation concerning their natures \cite{Stancu:2009ka,Wang:2015pea,Wang:2016tzr,Wang:2016gxp,Chen:2016oma}.

Concerning those exotic states, apart from the genuine resonances explanations, such as molecular states, tetraquark states or hybrid, some non-resonance explanations were also proposed in literatures.
There has been many theoretical attempts to try to connect the singularities of the rescattering processes with the resonance-like peaks in experiments, such as the cusp effect~\cite{Chen:2011pv,Chen:2011zv,Bugg:2011jr,Chen:2011xk,Swanson:2015bsa,Swanson:2014tra}, or the triangle singularity mechanism\cite{Wu:2011yx,Ketzer:2015tqa,Wang:2013cya,Liu:2013vfa,Liu:2014spa,Szczepaniak:2015eza,Guo:2014iya,Liu:2015taa,Liu:2015cah,Liu:2016dli,Liu:2015fea,Guo:2015umn,Aceti:2012dj,Achasov:2015uua}.
It is shown that sometimes it is not necessary to introduce a genuine resonance to describe a resonance-like peak, because some kinematic singularities of the rescattering amplitudes could behave themselves as bumps in the corresponding invariant mass distributions, which may bring ambiguities to our understanding about the nature of exotic states. 
Before claiming that one resonance-like peak corresponds to one genuine particle, it is also necessary to exclude or confirm these possibilities.

In this work, we investigate the possible rescattering effects in the process $B^+\to J/\psi\phi K^+$.
The open-charmed mesons rescatterings and $\psi^\prime\phi$ rescatterings are studied in Section II. The conclusions and some discussions are given in Section III.


\begin{figure}[htb]
	\centering
	\includegraphics[width=0.7\hsize]{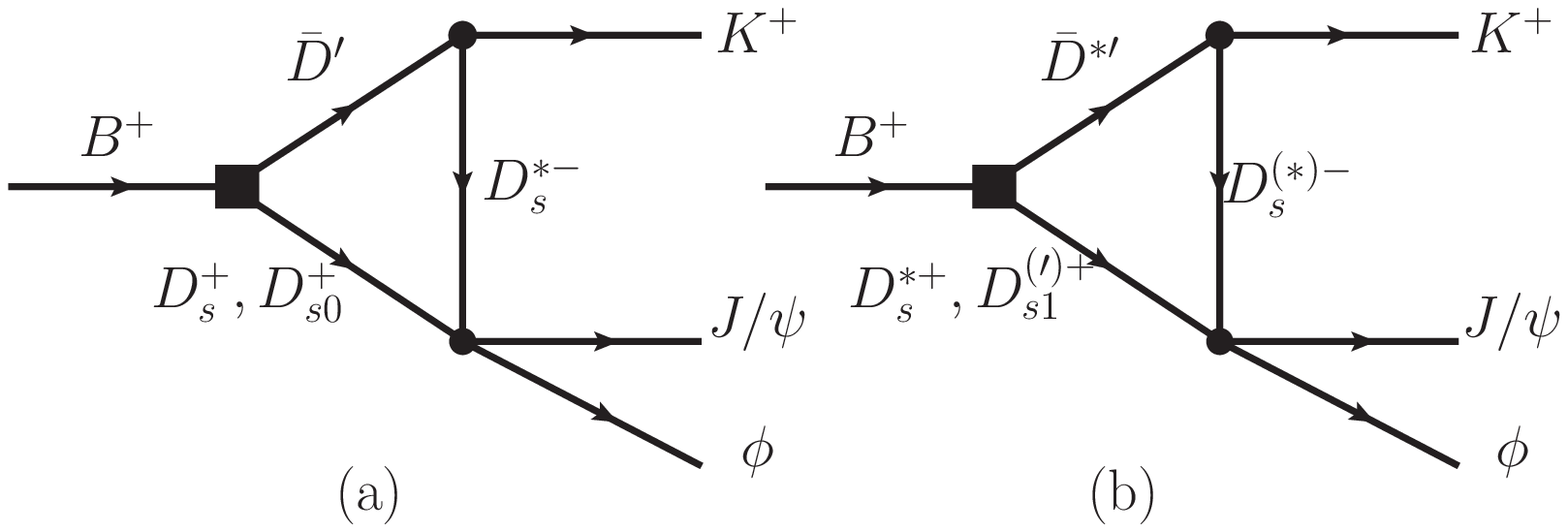}
	\caption{Triangle rescattering diagram via the open-charmed meson loops.}\label{Diagram-opencharm}
\end{figure}

\section{Rescattering Effects}
\subsection{Open-charmed mesons rescattering effect}
In experiments, the rate of $B$ decays into a charmed meson and a charmed-strange meson is found to be quite large among the hadronic decay modes. Since the velocities of the open-charmed mesons will not be large, they have adequate time to get involved in the final state rescatterings. It has been suggested that the rescattering effect may play an important role in the hadronic decays of bottom meson \cite{Colangelo:2002mj,Colangelo:2003sa,Cheng:2004ru,Wang:2008hq,Xu:2016kbn}. For higher excited charmed mesons, they may further decay into a kaon and a charmed-strange meson, and then we may expect that the rescattering processes illustrated in Fig.\ref{Diagram-opencharm} would contribute to the decay channel $B^+\to K^+ J/\psi \phi$. The lowest charmed mesons which can decay into $D_{s}^{(*)}K$ could be the first radially excited states $D^{\prime}$ and $D^{*\prime}$. The experimentally observed resonances corresponding to $D^{\prime}$ and $D^{*\prime}$ are $D(2550)$ and $D(2600)$ respectively \cite{Agashe:2014kda}, which is widely accepted. According to the latest results of LHCb \cite{Aaij:2013sza}, their masses and widths are
\begin{eqnarray}\label{Eq-Mass-D}
&&	M_{D^\prime}=2579.5\pm 3.4\pm 5.5\ \mbox{MeV},\ 	\Gamma_{D^\prime}=177.5\pm 17.8\pm 46.0\ \mbox{MeV},  \nonumber \\
&& M_{D^{*\prime}}=2649.2\pm 3.5\pm 3.5\ \mbox{MeV},\ 	\Gamma_{D^{*\prime}}=140.2\pm 17.1\pm 18.6\ \mbox{MeV}. 
\end{eqnarray}  
Since the mass of $D^\prime$ is somewhat lower than the threshold of $D_s^* K$ ($\sim 2606$ MeV), its contribution to the rescattering amplitude is supposed to be smaller compared with $D^{*\prime}$. There exist theoretical and experimental indications that the rate of $B$ decays into excited charmed mesons would be sizable, although such decays are supposed to be suppressed by the heavy quark symmetry (HQS) at the leading order \cite{Ebert:1999ed,delAmoSanchez:2010vq,Bernlochner:2012bc,Segovia:2013sxa,Becirevic:2013mp}. This implies that the rescattering effects induced by those excited states may be important.

Another interesting property of Fig.\ref{Diagram-opencharm} is that the thresholds of some $D_{sJ}^{(*)+}D_s^{(*)-}$ combinations  are rather close to the ``$X$'' states observed in $J/\psi\phi$ distributions. For convenience, we use $D_{s0}$, $D_{s1}$ and $D_{s1}^\prime$ to represent the $P$-wave charmed-strange mesons $D_{s0}(2317)$, $D_{s1}(2460)$ and $D_{s1}(2536)$, respectively. From Table.~\ref{thresholdtab}, one can see that the thresholds of $D_{s}^{*} D_s$, $D_{s}^{*} D_{s}^{*}$ ($D_{s0}D_{s}$), and $D_{s1}^\prime D_s$ ($D_{s1}D_s$, $D_{s1}D_{s}^{*}$, $D_{s0}D_{s}^{*}$) are close to the masses of $X(4140)$, $X(4274)$ and $X(4500)$, respectively. Due to the singularities around these thresholds may be present in the rescattering amplitudes. One may wonder whether there are some connections between the singularities and the ``$X$'' states. There are two intriguing singularities which may appear in the triangle rescattering diagrams. When two of the three intermediate states are on-shell, the singularity at threshold is a finite square-root branch point, which corresponds to the cusp effect. When all of the three intermediate states can be on-shell simultaneously, there will be a triangle singularity in the amplitude, which may result in narrow peaks in the corresponding spectrum \cite{Liu:2015taa}.    

\begin{table}\caption{Thresholds for the ${D}_{sJ}^{(*)+} D_s^{(*)-}$.} 
	\begin{center}
		\begin{tabular}{|c|c|c|c|c|c|c}
			\hline
	\text{Threshold [MeV]} & $D_s^+$ & $D_s^{*+}$ & $D_{s0}^{+}(2317)$ & $D_{s1}^{+}(2460)$ & $D_{s1}^{+}(2536)$ \\
	\hline
	$D_s^-$ & $ 3936.6 $ & $ 4080.4 $ & $ 4286.0 $ &$  4427.8 $ & $ 4503.4 $ \\
	\hline
	$D_s^{*-}$ & $ 4080.4 $ & $ 4224.2 $ & $ 4429.8 $ & $ 4571.6 $ & $ 4647.2 $ \\
	\hline
		\end{tabular}
	\end{center}\label{thresholdtab}
\end{table} 



\subsubsection{The Model}
In the factorization approach,  if the contributions from penguin operators are neglected, the decays $B\to D_{sJ}^{(*)} \bar{D}^{(*)\prime} $ receive contributions only from the external $W$-emission diagram. The weak amplitude $\left\langle D_{sJ}^{(*)} \bar{D}^{(*)\prime}| H_W|B \right\rangle $ can then be factorized into the product of two matrix elements, i.e.,
\begin{eqnarray}\label{weakamp}
\left\langle D_{sJ}^{(*)} \bar{D}^{(*)\prime}| H_W|B \right\rangle= \frac{G_F}{\sqrt{2}}  V_{cb}^{*} V_{cs}  a_1 \left\langle  \bar{D}^{(*)\prime}| (V-A)^\mu|B \right\rangle \left\langle D_{sJ}^{(*)}| (V-A)_\mu|0 \right\rangle,
\end{eqnarray}
with the Wilson-coefficient combination $a_1=c_1+c_2/N_c$. In the framework of heavy quark effective theory, the matrix element $\left\langle  \bar{D}^{(*)\prime}| (V-A)^\mu|B \right\rangle$ is parametrized by a series of hadronic form factors:
\begin{eqnarray}\label{formfactor}
	\left\langle  \bar{D}^{\prime}| V^\mu|B \right\rangle &=& \sqrt{M_B M_{D^\prime}} \left[ h_+ (\omega) (v+v^\prime)^\mu  + h_- (\omega) (v-v^\prime)^\mu  \right], \nonumber \\
	\left\langle  \bar{D}^{\prime}| A^\mu|B \right\rangle &=& 0,\nonumber \\
	\left\langle  \bar{D}^{*\prime}| V^\mu|B \right\rangle &=& \sqrt{M_B M_{D^{*\prime}}} \left[ i h_V(\omega) \varepsilon^{\mu\nu\alpha\beta}\epsilon_{\nu}^* v^\prime_\beta v_\alpha  \right],\nonumber \\
	\left\langle  \bar{D}^{*\prime}| A^\mu|B \right\rangle &=& \sqrt{M_B M_{D^{*\prime}}} \left[ h_{A_1}(\omega) (\omega+1) \epsilon^{*\mu} -(h_{A_2}(\omega) v^\mu +h_{A_3}(\omega) v^{\prime\mu} ) (\epsilon^* \cdot v) \right],
\end{eqnarray}
where $v$ ($v^\prime$) is the velocity of $B$ ($\bar{D}^{(*)\prime}$), $\omega$ is the product of velocities $v\cdot v^\prime$, and $\epsilon$ is the polarization vector of $\bar{D}^{*\prime}$. For the decay process $B\to D_{sJ}^{(*)} \bar{D}^{(*)\prime} $, in the rest frame of $B$, both $D_{sJ}^{(*)}$ and $\bar{D}^{(*)\prime}$ nearly stay at rest, which is very close to the zero recoil limit $\omega=1$. As an approximation, we will set $v=v^\prime=(1,0,0,0)$ in the following sections, and calculate the numerical results at the zero recoil limit. In Eq.~(\ref{weakamp}), the matrix element $\left\langle D_{sJ}^{(*)}| (V-A)_\mu|0 \right\rangle$ is related with the decay constant of the corresponding $ D_{sJ}^{(*)}$, which is defined as
\begin{eqnarray}
	\left\langle D_s| A_\mu|0 \right\rangle &=& f_{D_s} M_{D_s} v_\mu, \nonumber \\
		\left\langle D_s^*| V_\mu|0 \right\rangle &=& f_{D_s^*} M_{D_s^*} \epsilon^*_\mu, \nonumber\\
			\left\langle D_{s0}| V_\mu|0 \right\rangle &=& f_{D_{s0}} M_{D_{s0}} v_\mu, \nonumber\\
\left\langle D_{s1}^{(\prime)}| A_\mu|0 \right\rangle &=& f_{D_{s1}^{(\prime)}} M_{D_{s1}^{(\prime)}} \epsilon^*_\mu.
\end{eqnarray}
At zero recoil, it can be noticed that in Eq.~(\ref{formfactor}) only the form factors $h_+$ and $h_{A_1}$  have the non-vanishing contributions. Correspondingly, for the rescattering processes $B^+\to J/\psi\phi K^+$ via the $D_{sJ}^{(*)} \bar{D}^{(*)\prime}$-loops, only the diagrams illustrated in Fig.~\ref{Diagram-opencharm} can survive. Explicitly, the sub-diagrams involved in the calculations are $D_{s}^{+} \bar{D}^{\prime 0} [D_{s}^{*-}]$, $D_{s0}^{+} \bar{D}^{\prime 0} [D_{s}^{*-}]$, $D_{s}^{*+} \bar{D}^{*\prime 0} [D_{s}^{-}]$, $D_{s}^{*+} \bar{D}^{*\prime 0} [D_{s}^{*-}]$, $D_{s1}^{+} \bar{D}^{*\prime 0} [D_{s}^{-}]$, $D_{s1}^{+} \bar{D}^{*\prime 0} [D_{s}^{*-}]$, $D_{s1}^{\prime+} \bar{D}^{*\prime 0} [D_{s}^{-}]$, and $D_{s1}^{\prime+} \bar{D}^{*\prime 0} [D_{s}^{*-}]$ loops, of which the particles in the brackets represent the exchanged mesons between $D_{sJ}^{(*)}$ and $\bar{D}^{(*)\prime}$.

In order to estimate the rescattering amplitudes, we also need to know the relevant strong couplings in Fig.~\ref{Diagram-opencharm}. To proceed, the momentum for external and internal  particles are denoted as $B^+(P)\to K^+(p_1) J/\psi(p_2)\phi(p_3)$ and $D_{sJ}^{(*)+}(q_3) \bar{D}^{(*)\prime 0} (q_2) [D_{s}^{(*)-}(q_1)] $, respectively. In the framework of heavy hadron chiral perturbation theory, the decay amplitudes for $\bar{D}^{(*)\prime}\to D_{s}^{(*)} K$ are given by 
\begin{eqnarray}\label{Dprimedecay}
\mathcal{A}(\bar{D}^{\prime}\to D_{s}^{*} K) &=& \frac{g}{f_K}\sqrt{M_{D^\prime}M_{D_s^*}}\  p_1\cdot\epsilon^{*}({D}_{s}^{*}), \nonumber \\
\mathcal{A}(\bar{D}^{*\prime}\to D_{s}^{} K) &=& \frac{g}{f_K}\sqrt{M_{D^{*\prime}}M_{D_s}}\  p_1\cdot \epsilon(\bar{D}^{*\prime}), \nonumber \\
\mathcal{A}(\bar{D}^{*\prime}\to D_{s}^{*} K) &=& i \frac{g}{f_K}\sqrt{M_{D^{*\prime}}M_{D_s^*}}\  \varepsilon^{\mu\nu\alpha\beta} p_{1\mu} v_{\nu} \epsilon_{\alpha}(\bar{D}^{*\prime}) \epsilon_{\beta}^{*}({D}_{s}^{*}) , 
\end{eqnarray}
where the relative coupling strength among different channels is determined by the HQS. For the $S$-wave scattering $D_s^{(*)+}D_s^{(*)-}\to J/\psi\phi$, we construct a contact interaction which respects the HQS in Ref. \cite{Liu:2015cah}, and the relevant scattering amplitude takes the form 
\begin{eqnarray}
\mathcal{A}(D_s^{*+}D_s^{-}\to J/\psi\phi) &=& i \beta_S \  \varepsilon^{\alpha\beta\gamma\delta} v_\alpha \epsilon_\beta(D_s^{*+}) \epsilon_\gamma^*(\phi) \epsilon_\delta^*(J/\psi), \nonumber \\
\mathcal{A}(D_s^{*+}D_s^{*-}\to J/\psi\phi) &=& \beta_S \  (-g^{\alpha\beta}g^{\gamma\delta}+g^{\alpha\delta}g^{\beta\gamma}+g^{\alpha\gamma}g^{\beta\delta}) \epsilon_\alpha(D_s^{*+}) \epsilon_\beta(D_s^{*-}) \epsilon_\gamma^*(\phi) \epsilon_\delta^*(J/\psi),
\end{eqnarray}
where $\beta_S$ is the coupling constant for the contact interaction. It should be mentioned that according to the above amplitudes, the rescattering amplitude corresponding to the  $D_{s}^{*+} \bar{D}^{*\prime 0} [D_{s}^{*-}]$-loop actually vanishes. This can be understood from the parity and angular momentum conservations. If $D_{s}^{*+}$ and $D_{s}^{*-}$ scatter in relative $S$-wave, their quantum numbers can only be $0^{++}$ or $2^{++}$, which means the quantum numbers of the produced $J/\psi\phi$ can only be $0^{++}$ or $2^{++}$. With an anti-symmetric tensor appearing in the rescattering amplitude, this sub-diagram finally gives a vanishing contribution. 

If the quantum numbers of $J/\psi\phi$ system are $J^{PC}=0^{++}$ or $J^{PC}=1^{++}$, the $P$-wave and $S$-wave charmed-strange mesons can scatter into $J/\psi\phi$ via relative $P$-wave. We assume the quantum numbers of $J/\psi\phi$ are $0^{++}$, then the $P$-wave scattering amplitudes which respect the HQS take the form
\begin{eqnarray}
 \mathcal{A}(D_{s0}^{+}D_s^{*-}\to J/\psi\phi) &=& \frac{\beta_P}{\sqrt{M_{D_{s0}}  M_{D_s^*}}}\ q_{D_{s0}} \cdot \epsilon(D_s^{*}) \  \epsilon^*(\phi)\cdot \epsilon^*(J/\psi) , \nonumber \\
  \mathcal{A}(D_{s1}^{(\prime)+}D_s^{-}\to J/\psi\phi) &=& \frac{\beta_P}{\sqrt{M_{D_{s1}^{(\prime)}}  M_{D_s}}}\ q_{D_s} \cdot \epsilon(D_{s1}^{(\prime)}) \  \epsilon^*(\phi)\cdot \epsilon^*(J/\psi) , \nonumber \\
\mathcal{A}(D_{s1}^{(\prime)+}D_s^{*-}\to J/\psi\phi) &=& \frac{i\beta_P}{{M_{D_{s1}^{(\prime)}}  M_{D_s^*}}}\ \varepsilon_{\mu\nu\alpha\beta}\  q_{D_s^*}^\alpha  q_{D_{s1}^{(\prime)}}^\beta  \epsilon^\mu(D_s^{*}) \epsilon^\nu(D_{s1}^{(\prime)})  \  \epsilon^*(\phi)\cdot \epsilon^*(J/\psi) ,
\end{eqnarray} 
where $\beta_P$ is the coupling constant. 

By means of the above scattering amplitudes, the rescattering amplitude of $B^+\to K^+ J/\psi \phi$ via the open charm loops in Fig.~\ref{Diagram-opencharm} is given by
\begin{eqnarray}\label{resamp}
	T_{B^+\to K^+ J/\psi \phi}^{[D_{sJ}^{(*)}\bar{D}^{(*)\prime}]}=\frac{1}{i} \int \frac{d^4q_1}{(2\pi)^4} \frac{\mathcal{A}(B\to D_{sJ}^{(*)} \bar{D}^{(*\prime)}) \mathcal{A}(\bar{D}^{(*)\prime}\to D_{s}^{(*)} K) \mathcal{A}(D_{sJ}^{(*)}D_s^{{*}}\to J/\psi\phi) }{(q_1^2-M_{D_s^{*}}^2) (q_2^2-M_{\bar{D}^{(*)\prime}}^2 +i M_{\bar{D}^{(*)\prime}}\Gamma_{\bar{D}^{(*)\prime}}) (q_3^2-M_{D_{sJ}^{(*)}}^2) },
\end{eqnarray}
where the sum over polarizations of intermediate states are implicit. As an approximation, we adopt a Breit-Wigner type propagator to account for the width effect of $\bar{D}^{(*)\prime}$ in the above loop integral.

\subsubsection{Numerical Results}
\begin{figure}[htb]
	\centering
	\includegraphics[width=8cm]{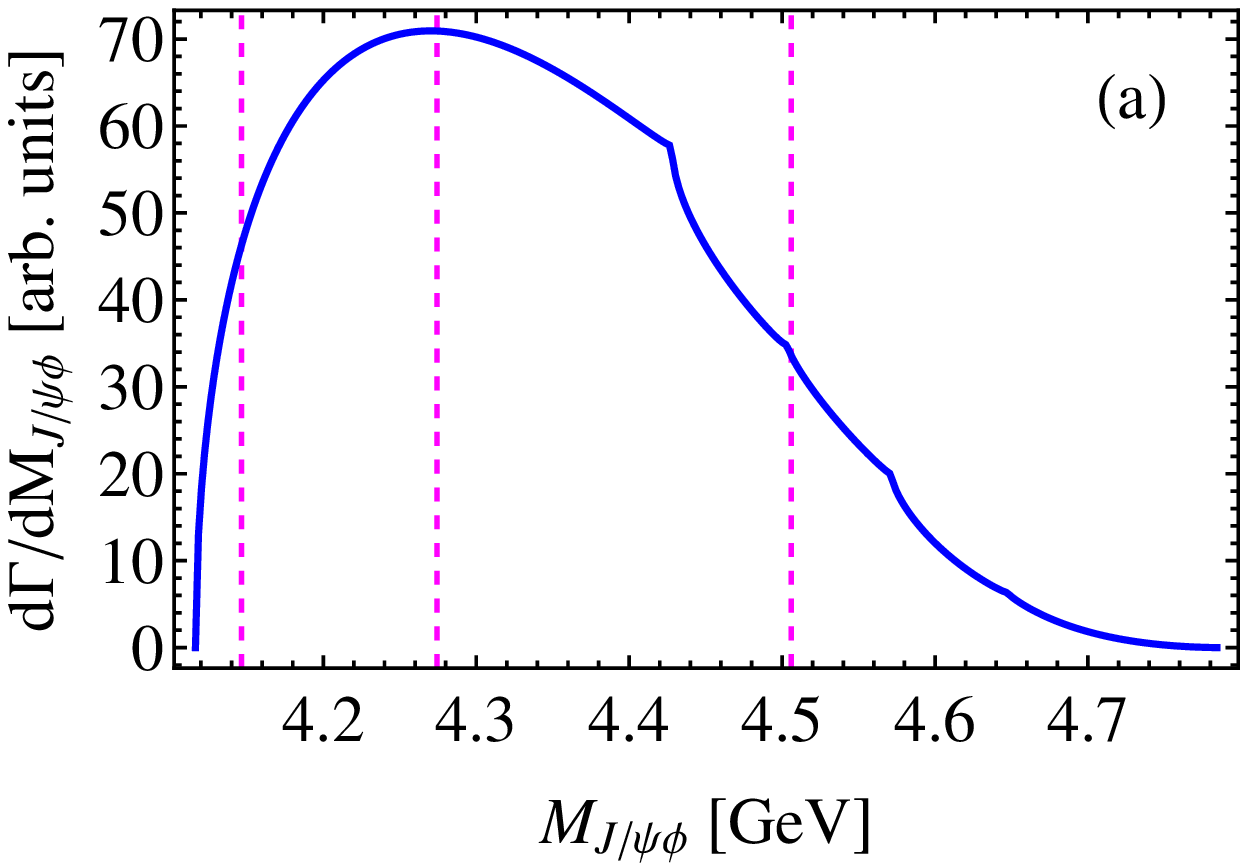}\\
	\includegraphics[width=8cm]{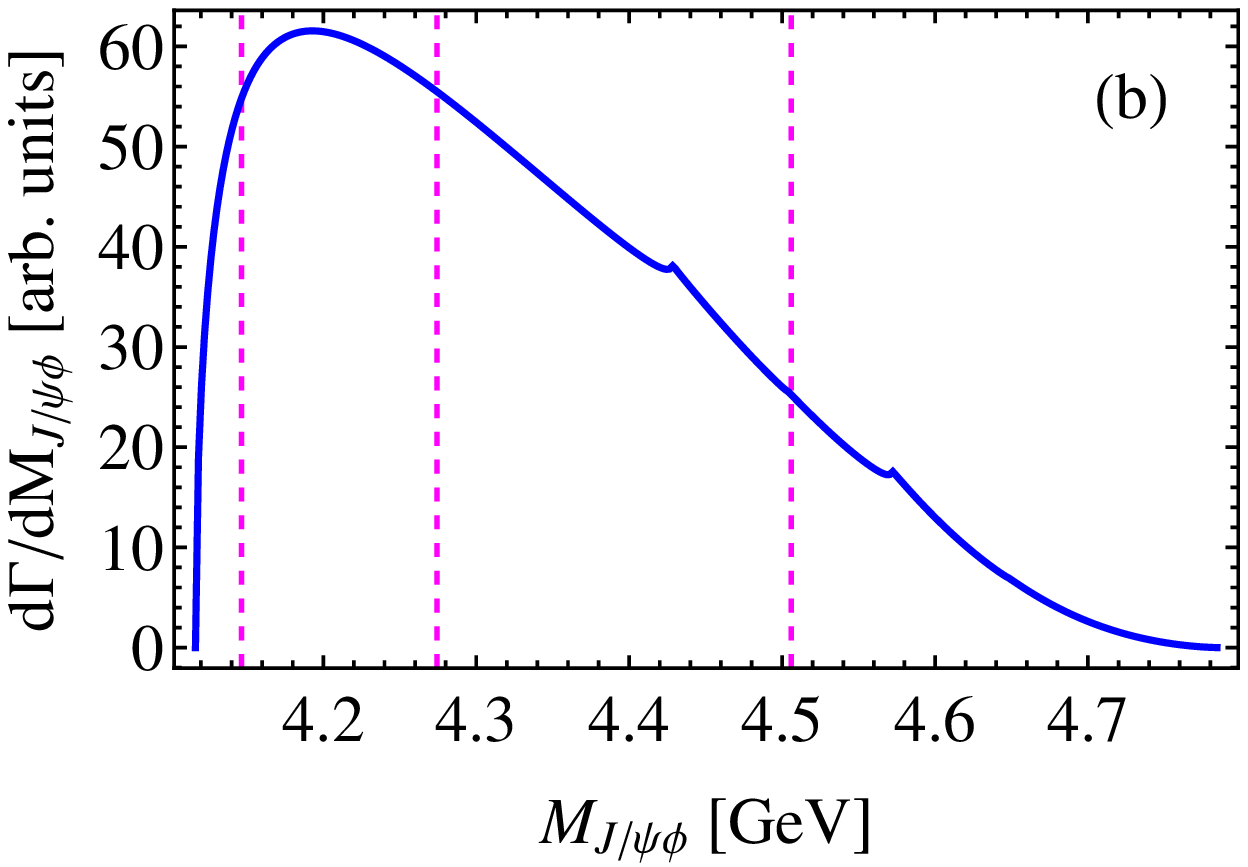}
	\caption{Invariant mass distributions of $J/\psi\phi$ via the rescattering processes of Fig.~\ref{Diagram-opencharm}. The mass (width) of $\bar{D}^{(*)\prime}$ is taken to be that of (a) the first  and (b) the second radially excited state of $\bar{D}^{(*)}$, respectively. The vertical dashed lines indicate the positions of $X(4140)$, $X(4274)$ and $X(4500)$ respectively. }\label{lineshape-Dloop}
\end{figure}

Ignoring the common coupling constant and form factors, the relative strength of different rescattering amplitudes mainly depend on the decay constants of $D_{sJ}^{(*)}$ and the form factors $h_+$ and $h_{A_1}$. For the decay constant of $D_s$, we adopt the experimental value, i.e., $f_{D_s}=257.5\ \mbox{MeV}$ \cite{Agashe:2014kda}, and make $f_{D_s^*}=f_{D_s}$. In the heavy quark limit, we have the following relations \cite{Isgur:1989vq,Isgur:1989ed,LeYaouanc:1996bd,Veseli:1996yg}
\begin{eqnarray}
	f_{D_{s0}}=f_{D_{s1}},\ f_{D_{s1}^\prime}=0. 
\end{eqnarray}
But these relations are not consistent with the experimental observations very well. In our calculation, we adopt the values calculated in the covariant light-front model \cite{Cheng:2003sm,Cheng:2006dm}, which gives
\begin{eqnarray}
	f_{D_{s0}}=71\ \mbox{MeV},\   f_{D_{s1}}=121\ \mbox{MeV},\ f_{D_{s1}^\prime}=38\ \mbox{MeV}.  
\end{eqnarray}
For the form factors $h_+$ and $h_{A_1}$, we adopt the values calculated in the framework of relativistic quark model \cite{Ebert:1999ed}, which gives
\begin{eqnarray}
	h_+(1)\simeq 0.012, \ h_{A_1}(1)\simeq 0.098.
\end{eqnarray}
Since $h_{A_1}(1)$ is much larger than $ h_+(1) $, correspondingly we can expect that the contribution of Fig.\ref{Diagram-opencharm}(b) would be much larger than Fig.\ref{Diagram-opencharm}(a). The numerical results of $J/\psi\phi$ invariant mass distributions via the rescattering processes are displayed in Fig.\ref{lineshape-Dloop}. The result in Fig.\ref{lineshape-Dloop}(a) is obtained by setting the values of $M_{\bar{D}^{(*)\prime}}$ and $\Gamma_{\bar{D}^{(*)\prime}}$ as those in Eq.~(\ref{Eq-Mass-D}).
 To check the dependence on the mass of intermediate states, we also calculate the rescattering amplitudes by setting the values of $M_{\bar{D}^{(*)\prime}}$ and $\Gamma_{\bar{D}^{(*)\prime}}$ as those of the second radially excited states, and the result is displayed in Fig.\ref{lineshape-Dloop}(b). There is no experimental measurement concerning the second radially excited charmed mesons $D^{(*)}(3S)$, and the following results calculated in the quark model are adopted in calculations \cite{Godfrey:2015dva}:   
\begin{eqnarray}\label{Eq-Mass-D3S}
&&	M_{D(3S)}=3068\ \mbox{MeV},\ 	\Gamma_{D(3S)}=106\ \mbox{MeV},  \nonumber \\
&& M_{D^{*}(3S)}=3110\ \mbox{MeV},\ 	\Gamma_{D^{*}(3S)}=103\ \mbox{MeV}. 
\end{eqnarray} 

In Figs.\ref{lineshape-Dloop}(a) and (b), one may notice that there are several cusps which stay around the thresholds of $D_{s1}D_s$, $D_{s1}^\prime D_s$, $D_{s1}D_{s}^{*}$ and $D_{s1}^\prime D_{s}^{*}$, respectively. As discussed previously, the sub-diagram $D_{s}^{*+} \bar{D}^{*\prime 0} [D_{s}^{*-}]$-loop gives vanishing contribution, therefore there is no cusp appearing around the $D_{s}^{*+}D_{s}^{*-}$ threshold.  Correspondingly, for the rescattering processes studied in this paper, there is no cusp that can simulate the structure of $X(4274)$.
Because the threshold of $J/\psi\phi$ is larger than that of $D_{s}^{*+}D_{s}^{-}$, there is no cusp corresponding to the $D_{s}^{*+}D_{s}^{-}$ threshold either. But an obvious threshold enhancement structure appears in the $J/\psi\phi$ distributions, which implies that the rescattering effect may simulate the structure of $X(4140)$. 
In Refs. \cite{Aaij:2016iza,Aaij:2016nsc}, by employing a cusp model proposed in Ref. \cite{Swanson:2015bsa}, the authors claimed that the $X(4140)$ may be described as a $D_{s}^{*\pm }D_{s}^{\mp}$ cusp, but a resonant interpretation is also possible.

All of the cusps around $X(4500)$ are too broad and small to simulate the structure of $X(4500)$. The is mainly because if the quantum numbers of $J/\psi\phi$ are set to be $0^{++}$, to preserve the parity, only via $P$-wave can $D_{s1}^{(\prime)}D_s^{(*)}$ scatter into $J/\psi\phi$. And usually the near threshold $P$-wave scatterings will be suppressed due to the small momentum of scattering particles. For the rescattering processes discussed here, the kinematic conditions for the presence of triangle singularities are not satisfied \cite{Liu:2015taa}. Therefore the rescattering amplitude cannot get enhancement induced by the triangle singularities either.

According to the above analysis, it seems hard to ascribe the observation of $X(4274)$ and $X(4500)$ in $B$ decays to the rescattering effects. This implies that $X(4274)$ and $X(4500)$ may correspond to some genuine resonances, such as tetraquark states or some higher excited charmonium states.

\subsection{$\psi^\prime \phi$ rescattering effect}

\begin{figure}[htb]
	\centering
	\includegraphics[width=0.4\hsize]{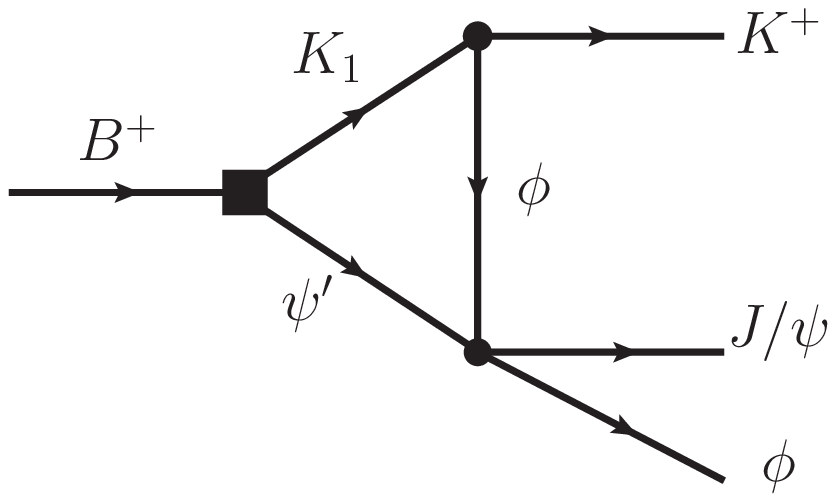}
	\caption{Triangle rescattering diagram via the $\psi^\prime K_1$ loop.}\label{Diagram-K1loop}
\end{figure}

\subsubsection{The Model}

In the naive factorization approach, the amplitude for the color suppressed decays $B\to M_{c\bar{c}} K^{(**)}$, with $M_{c\bar{c}}$ and $K^{(**)}$ representing a charmonia and a kaon ( excited kaon) meson respectively, is given by 
\begin{eqnarray}\label{weakpsiK}
\left\langle M_{c\bar{c}}  K^{(**)+} | H_W|B^+ \right\rangle= \frac{G_F}{\sqrt{2}} V_{cb}^{*} V_{cs}   a_2 \left\langle  K^{(**)+} |(V-A)^\mu|B^+ \right\rangle \left\langle M_{c\bar{c}} | (V-A)_\mu|0 \right\rangle,
\end{eqnarray}
with $a_2=c_2+c_1/N_c$. The matrix element $\left\langle M_{c\bar{c}} | (V-A)_\mu|0 \right\rangle$ is proportional to the decay constant of $M_{c\bar{c}}$. It can be expected that the ratio of  the amplitude $|\mathcal{A}(B\to J/\psi K^{**}) |$ to the amplitude $ |\mathcal{A}(B\to \psi^\prime K^{**})|$ will be close to $f_{J/\psi}/f_{\psi^\prime}$.  For $J/\psi$ and $\psi^\prime$, the discrepancy between the decay constants $f_{J/\psi}$ ($416\pm 5\ \mbox{MeV}$) and $f_{\psi^\prime}$ ($294\pm 5\ \mbox{MeV}$) is not very large. 
In the analysis of LHCb concerning the decay $B^+\to K^+ J/\psi \phi$, it is shown that there is a rich spectrum of excited kaon resonances, which has significant contributions via the sequential decays $B^+\to J/\psi K^{**+} \to J/\psi \phi K^+$.
Although the reflection of excited kaons can not result in obvious resonance-like structures in $J/\psi\phi$ distributions, they may contribute to the production of those ``$X$" states in another way. As discussed above, it can be expected that the decay rate of $B\to \psi^\prime K^{**}$ would be at the same order of magnitude with that of $B\to J/\psi K^{**}$. For some higher excited kaons, their on-shell production may be prohibited due to the phase space, but taking into account their broad decay widths, they can still contribute to the process $B^+\to \psi^\prime \phi K^+$.

Interestingly, it is found that the threshold of $\psi^\prime\phi$ ($\sim 4706\ \mbox{MeV}$) is very close to the mass of $X(4700)$  ($\sim 4704\ \mbox{MeV}$). Therefore it is possible that the rescattering process illustrated in Fig.~\ref{Diagram-K1loop} may result in some  resonance-like peaks in $J/\psi\phi$ distributions around the $\psi^\prime\phi$ threshold, which may simulate the $X(4700)$ signal.
Among the excited kaons, the dominant contributions come from the axial vector states, as shown in Refs. \cite{Aaij:2016iza,Aaij:2016nsc}. In the following analysis, we will focus on the rescattering effects induced by $K_1$, of which the quantum numbers are $J^P=1^+$.

The general invariant amplitude for $B\to \psi^\prime K_1$ can be written as:
\begin{eqnarray}
	\mathcal{A}(B\to \psi^\prime K_1) &=& a\ \epsilon^*(\psi^\prime) \cdot \epsilon^*(K_1) + \frac{b}{(M_B +M_{K_1})^2}\  p_B \cdot \epsilon^*(\psi^\prime) \ p_B \cdot \epsilon^*(K_1) \nonumber\\
	&+& \frac{c}{(M_B +M_{K_1})^2}\ i\varepsilon_{\mu\nu\alpha\beta} p_B^\mu p_{K_1}^\nu \epsilon^{*\alpha}(\psi^\prime) \epsilon^{*\beta}(K_1).
\end{eqnarray}
For $B$ decaying into  $\psi^\prime$ and the higher excited state $K_1$, both $\psi^\prime$ and $K_1$ will nearly stay at rest in the rest frame of $B$. Therefore in the above equation, only the first term on the right hand side will contribute significantly. As an approximation, we only keep the fist term in the calculation, and set the form factor $a$ as a constant. The axial-vector meson $K_1$ can decay into $\phi K$ in $S$-wave, and the amplitude takes the form
\begin{eqnarray}
	\mathcal{A}(K_1\to \phi K) = g_{K_1}  \epsilon(K_1) \cdot \epsilon^*(\phi),
\end{eqnarray}
where $g_{K1}$ is the coupling constant. $\psi^\prime\phi$ can scatter into $J/\psi\phi$ via exchanging soft gluons. To simplify the model, we only construct a contact interaction for this scattering, and the amplitude read
\begin{eqnarray}
	\mathcal{A}(\psi^\prime\phi \to J/\psi\phi )= g_{CT}\ \epsilon(\psi^\prime) \cdot \epsilon(\phi) \ \epsilon^*(J/\psi) \cdot \epsilon^*(\phi),
\end{eqnarray}
where we have assumed the quantum numbers of $J/\psi\phi$ system are $0^{++}$, to be consistent with the quantum numbers of $X(4700)$.

The rescattering amplitude of $B^+\to K^+ J/\psi \phi$ via the $\psi^\prime K_1$-loop in Fig.~\ref{Diagram-K1loop} is given by
\begin{eqnarray}\label{psipK1amp}
T_{B^+\to K^+ J/\psi \phi}^{[\psi^\prime K_1]}=\frac{1}{i} \int \frac{d^4q_1}{(2\pi)^4} \frac{\mathcal{A}(B\to \psi^\prime K_1) \mathcal{A}(K_1\to \phi K) \mathcal{A}(\psi^\prime\phi \to J/\psi\phi ) }{(q_1^2-M_{\phi}^2) (q_2^2-M_{K_1}^2 +i M_{K_1}\Gamma_{K_1}) (q_3^2-M_{\psi^\prime}^2) },
\end{eqnarray}
where the sum over polarizations of intermediate states are implicit.

\subsubsection{Numerical Results}

\begin{figure}[htb]
	\centering
	\includegraphics[width=9cm]{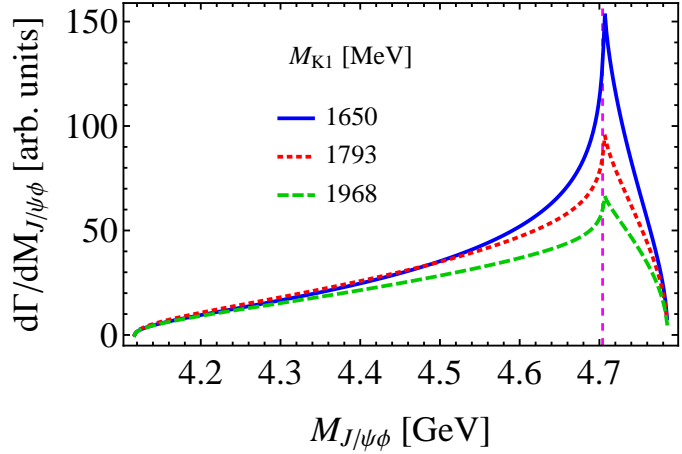}\\
	\caption{Invariant mass distributions of $J/\psi\phi$ via the rescattering processes of Fig.~\ref{Diagram-K1loop}. The vertical dashed line indicate the position of  $X(4700)$. }\label{lineshape-K1loop}
\end{figure}

For the moment, we just focus on the lineshape behavior of the $J/\psi\phi$ distributions via the rescattering processes, but ignore the explicit values of the relevant coupling constants and form factors. The numerical results are displayed in Fig.~\ref{lineshape-K1loop}. According to the fitting results of LHCb \cite{Aaij:2016iza,Aaij:2016nsc}, the distributions in Fig.~\ref{lineshape-K1loop} are obtained by setting the mass (width) of $K_1$ to be $1650\ \mbox{MeV}$ ($150\ \mbox{MeV}$), $1793\ \mbox{MeV}$ ($365\ \mbox{MeV}$) and $1968\ \mbox{MeV}$ ($396\ \mbox{MeV}$) separately. From Fig.~\ref{lineshape-K1loop}, one can see that there is a clear peak stays around $4.7$ GeV, which corresponds to the $\psi^\prime\phi$ threshold. The $S$-wave near threshold scattering make this peaks obvious. Furthermore, although the kinematic condition for the triangle singularity in the rescattering process of Fig. \ref{Diagram-K1loop} is not fulfilled well either, it is already very close to the kinematic region where the triangle singularity can be present \cite{Liu:2015taa}. Therefore the physical rescattering amplitude will be influenced by the triangle singularity to some extent.

\begin{figure}[htb]
	\centering
	\includegraphics[width=9cm]{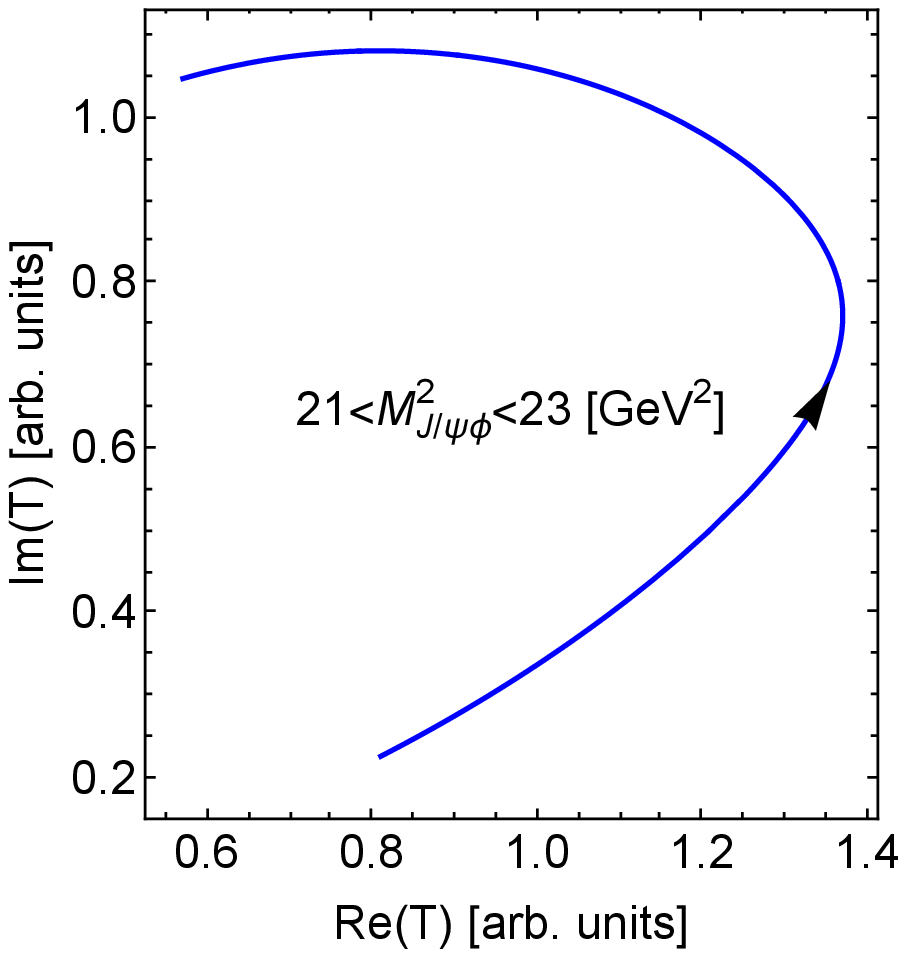}\\
	\caption{Real and imaginary parts of the rescattering amplitude in Eq.~(\ref{psipK1amp}). Motion with the increasing invariant mass $M_{J/\psi\phi}$ is counter-clockwise.  }\label{Argand}
\end{figure}

The Argand plot corresponds to the rescattering amplitude in Eq.~(\ref{psipK1amp}) is displayed in Fig.~\ref{Argand}. It can be seen that the phase of the amplitude shows a behavior of rapid counter-clockwise change, which is similar with a genuine resonance.


\section{Conclusion and Discussions}
In conclusion, it is possible that the $D_{s}^{*+}D_{s}^{-}$ rescattering via the open-charmed meson loops, and $\psi^\prime \phi$ rescattering via the $\psi^\prime K_1$ loops may simulate the structures of $X(4140)$ and $X(4700)$, respectively. However, if the quantum numbers of $X(4274)$ ($X(4500)$) are $1^{++}$ ($0^{++}$), due to the parity and angular momentum conservation, and the near threshold $P$-wave scattering characteristic, it is hard to describe the structures of $X(4274)$ and $X(4500)$ with rescattering effects, which implies that $X(4274)$ and $X(4500)$ could be genuine resonances. 

Concerning $X(4274)$, although it is observed in the $J/\psi\phi$ invariant mass distributions, it does not necessarily mean that it contains $s\bar{s}$ as the valence quarks. We suggest that it may be the conventional orbitally excited state $\chi_{c1}(3P)$. This suggestion is based on the following arguments:\\
\\
Firstly, the predicted mass of $\chi_{c1}(3P)$ in the framework of quark models is about $4271\ \mbox{MeV}$ or $4317\ \mbox{MeV}$ \cite{Barnes:2005pb,Godfrey:1985xj}, which is very close to the mass of $X(4274)$. The predicted width ($\sim 39$ MeV) is also close to the observed width of $X(4274)$ ($\sim 56$ MeV).  Although for the higher charmonium states, the prediction of conventional quark models is not very reliable, it can still be taken as a guidance;
\\
Secondly, if the $X(4274)$ is $\chi_{c1}(3P)$, apart from $J/\psi \phi$, one may expect that it can also easily decay into $J/\psi\omega$. From the Fig.2 in Ref. \cite{delAmoSanchez:2010jr}, it can be noticed that apart from $X(3872)$, there are also some bumps around $4.3$ GeV appearing in the $J/\psi\omega$ distributions.
Although the current statics for $B\to J/\psi \omega K$ may not be large enough to confirm this, it can still be taken as an evidence;
\\
Thirdly, the rescattering effects can not describe the observation of $X(4274)$, as discussed in this paper and Refs. \cite{Aaij:2016iza,Aaij:2016nsc};
\\
Fourthly, under the factorization ansatz, since $\chi_{c0}$, $\chi_{c2}$ and their radially excited states can not be produced via the $V-A$ current, it can be expected that the decay rate of $B\to K\chi_{c1}(3P)$ will be larger than that of $B\to K\chi_{c0}(3P)$ or $B\to K\chi_{c2}(3P)$. This may lead to that in the $J/\psi\phi$ distributions, only the $\chi_{c1}(3P)$ signal is significant.
\\

If the $X(4274)$ is really $\chi_{c1}(3P)$, one may search for it in the radiative decays of $\psi(4415)$ ($\psi(4S)$), taking into account that this $E1$ transition is predicted to have a significantly large branching ratios in the framework of quark model \cite{Barnes:2005pb,Godfrey:1985xj}.




\subsection*{Acknowledgments}
Helpful discussions with Qiang Zhao and Gang Li are acknowledged.
This work is supported in part by the Japan Society for the Promotion of Science under Contract No. P14324, and the JSPS KAKENHI (Grant No. 25247036).

\end{document}